\documentclass{article}

\usepackage{arxiv}
\usepackage[utf8]{inputenc} % allow utf-8 input
\usepackage[T1]{fontenc}    % use 8-bit T1 fonts
\usepackage{hyperref}       % hyperlinks
\usepackage{url}            % simple URL typesetting
\usepackage{booktabs}       % professional-quality tables
\usepackage{amsmath}
\usepackage{amsfonts}       % blackboard math symbols
\usepackage{nicefrac}       % compact symbols for 1/2, etc.
\usepackage{lipsum}
\usepackage{graphicx}
\graphicspath{ {./images/} }

\title{\textbf{Hierarchical Decoupling of Informational and Energetic Channels in Dissipative Chaos: From Microscopic Tangles to Macroscopic Transport}}

\author{
    Takashi Hikihara\textsuperscript{1,2} \\
    \small{\textsuperscript{1} Kyoto University, Kyoto, Japan} \\
    \small{\textsuperscript{2} Research Center for Open Science and Data Platform,} \\
    \small{National Institute of Informatics, Tokyo, Japan}
}

\begin{document}
\maketitle
\begin{abstract}
Characterizing deterministic chaos in driven dissipative systems presents a structural challenge: abstract geometric frameworks identify topological horseshoes ($h_T > 0$) without certifying their asymptotic physical measures, while applied engineering paradigms rely on coarse-grained macroscopic observables (e.g., Lyapunov exponents, steady-state variance) that project multi-scale manifold dynamics into low-dimensional indicators. In this study, we demonstrate that phase-space volume contraction structurally decouples the informational channel, governed by topological complexity and Kolmogorov--Sinai entropy ($h_{KS}$), from the energetic channel governing macroscopic dissipation, power balance, and transport. We formulate a unified three-tier hierarchical architecture bridging: (I) microscopic topology (invariant manifold intersections), (II) ergodic physical measures (metric entropy and Lyapunov spectra), and (III) macroscopic ensemble statistics (variance and transport flux). Applying high-throughput deterministic binarization and saddle-continuation algorithms to the forced two-well Duffing oscillator reveals the origin of classical predictive limitations. Specifically, dense homoclinic tangles frequently degenerate into non-retaining transient saddles ($h_T > 0, h_{KS} = 0$) where energy flows into periodic sinks, whereas macroscopic chaos can persist beyond the destruction of primary homoclinic tangencies through higher-order saddle relays. We further generalize this hierarchical framework to periodic-potential rotational systems, showing how global heteroclinic crossings and resonant Shapiro steps enforce geometric bounds on macroscopic phase slips and rotational transport. By structurally disentangling the geometric capacity for information from macroscopic energetic realization, this framework links geometric mechanics with non-equilibrium statistical observability, providing a new diagnostic capability to predict noise-induced instabilities and unexplained chaotic regimes in applied physical systems.
\end{abstract}


\section{Introduction}

The transversal intersection of stable and unstable invariant manifolds is a fundamental geometric mechanism generating deterministic chaos and complex transport in nonlinear dynamical systems \cite{Smale1967, Guckenheimer1983}. Classically, analytical thresholds such as the Melnikov method provide criteria for the onset of homoclinic tangencies, delineating the theoretical boundaries of topological chaos \cite{Melnikov1963}. According to the Smale-Birkhoff theorem, the existence of a transversal homoclinic intersection implies the formation of a Smale horseshoe and a positive topological entropy ($h_T > 0$). However, in the context of driven dissipative systems, a discrepancy remains between this microscopic geometric threshold and macroscopically observable chaos. Numerical and physical experiments demonstrate that dissipative trajectories frequently collapse into periodic windows or phase-locked states (Arnold tongues) far outside the boundaries predicted by geometric criteria \cite{Ueda1987}.

Historically, this discrepancy has led to a conceptual divide in the study of non-equilibrium dynamics. Abstract geometric and topological approaches frequently identify horseshoe structures ($h_T > 0$) without determining whether these structures possess asymptotic physical measures. Conversely, applied perspectives rely on coarse-grained macroscopic observables---such as maximal Lyapunov exponents ($\lambda_1$) and steady-state kinetic variance---which project high-dimensional manifold causality into low-dimensional indicators. We argue that the root of this divide lies in the fundamental nature of phase-space volume contraction in dissipative systems. Specifically, dissipation structurally decouples the \textit{informational channel}, governed by topological complexity and Kolmogorov-Sinai metric entropy ($h_{KS}$), from the \textit{energetic channel}, which governs macroscopic power balance, kinetic fluctuations, and transport.

\begin{figure}[htbp]
    \centering
    \includegraphics[width=1.0\textwidth]{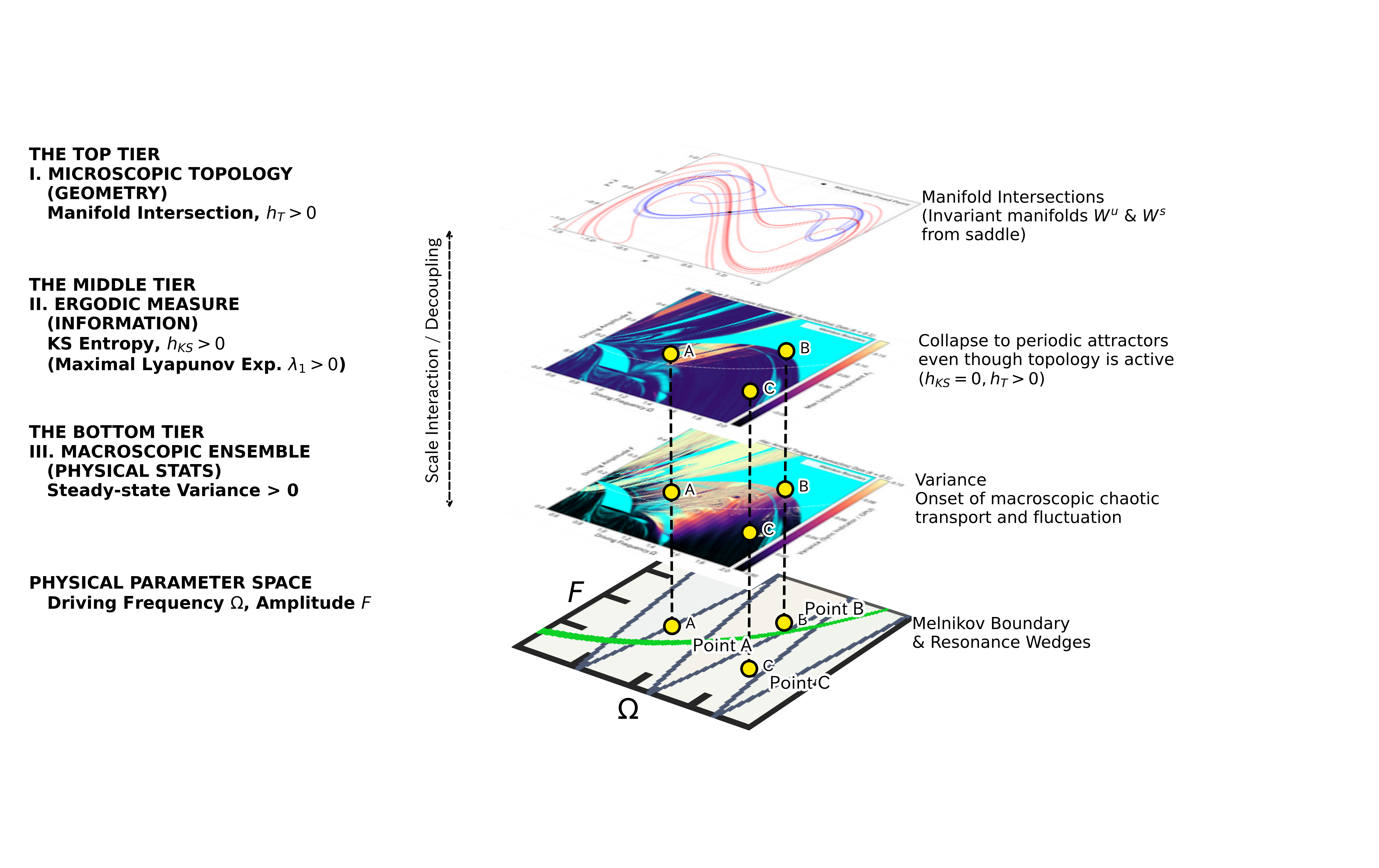}
    \caption{\textbf{Three-tier hierarchical architecture and structural decoupling in driven dissipative dynamical systems.} 
    \textbf{(a) Physical Parameter Space:} The base plane defined by driving frequency $\Omega$ and amplitude $F$, featuring the analytical Melnikov threshold curve (solid green curve) and subharmonic resonance wedges (Arnold tongues). Representative parameter regimes are indicated: Point A (sustained ergodic chaos), Point B (transient chaotic saddle), and Point C (higher-order saddle-driven chaos). 
    \textbf{(b) Bottom Tier (III. Macroscopic Ensemble):} Observable steady-state physical statistics (kinetic variance and macroscopic transport flux). While this tier reflects the thermodynamic energy balance between external power injection and friction, it fails to distinguish between laminar sinks and non-retained chaotic repellers. 
    \textbf{(c) Middle Tier (II. Ergodic Measure):} Asymptotic physical measure and informational production rate quantified by the Kolmogorov--Sinai entropy ($h_{KS} > 0$) and the maximal Lyapunov exponent ($\lambda_1 > 0$). Entrainment into periodic sinks collapses this tier to zero despite active microscopic tangles. 
    \textbf{(d) Top Tier (I. Microscopic Topology):} Microscopic geometric skeleton governed by the transverse intersection of stable ($W^s$, red) and unstable ($W^u$, blue) invariant manifolds emanating from saddle fixed points, establishing positive topological entropy ($h_T > 0$). 
    Vertical dashed trajectories crossing all tiers demonstrate the dynamic decoupling: Point B exhibits active microscopic geometry ($h_T > 0$) coexisting with an informational collapse ($h_{KS} = 0$), whereas Point C sustains macroscopic chaotic transport ($h_{KS} > 0, \text{Variance} > 0$) via higher-order saddle cascades even after the primary homoclinic tangency is severed.}
    \label{fig:concept}
\end{figure}

To evaluate this structural disconnect, this paper proposes a unified three-tier hierarchical architecture that explicitly separates these dynamic layers (see Fig.~\ref{fig:concept}). The framework comprises: (I) the \textit{microscopic topology} tier, representing the geometric capacity for chaos via manifold intersections ($h_T$); (II) the \textit{ergodic measure} tier, quantifying the asymptotic informational production rate via metric entropy ($h_{KS}$) and Lyapunov spectra; and (III) the \textit{macroscopic ensemble} tier, capturing observable physical statistics such as variance and phase-space transport flux. By evaluating dynamical states vertically across these three distinct tiers, we identify the mechanisms of predictive failure in single-layer indicators. For instance, parameter regimes exhibiting active microscopic tangles ($h_T > 0$) routinely experience an informational collapse ($h_{KS} = 0$) where energy flows into periodic sinks, forming non-retaining transient chaotic saddles. Conversely, macroscopic chaotic fluctuations can persist beyond the destruction of primary homoclinic tangencies through the relay of higher-order saddles.

The remainder of this paper is organized to demonstrate the universality of this hierarchical decoupling from local oscillations to global transport. In Sec.~\ref{sec:architecture}, we define the three-tier framework. In Sec.~\ref{sec:duffing}, we apply high-throughput, non-interpolated GPU-accelerated binarization and spectral analysis to the forced two-well Duffing oscillator, charting the boundaries where macroscopic ergodicity breaks away from microscopic topology. In Sec.~\ref{sec:rotational}, we generalize this framework to driven rotational systems with periodic potentials (pendulum-type dynamics), revealing how the transition from localized homoclinic tangles to global heteroclinic networks dictates macroscopic rotational transport. Discussion and conclusions are presented in Sec.~\ref{sec:discussion} and Sec.~\ref{sec:conclusion}.

%%%%%% Figure 1 %%%%%%%%

\section{The Three-Tier Hierarchical Architecture}
\label{sec:architecture}

To evaluate the structural decoupling between informational and energetic channels, we construct a unified hierarchical architecture applicable to general driven dissipative systems. As a representative testbed, we consider the forced two-well Duffing oscillator, which provides a canonical description of bistable localized dynamics:
\begin{equation}
    \ddot{x} + k\dot{x} - x + x^3 = F \cos(\Omega t), \label{eq:duffing}
\end{equation}
where $k$ represents the linear viscous damping coefficient, and $F$ and $\Omega$ denote the amplitude and frequency of the external harmonic forcing, respectively. In the first part of this study, the dissipation is fixed at $k=0.1$, allowing us to scan the two-dimensional physical parameter space spanned by $(\Omega, F)$. The continuous-time flow is discretized via the stroboscopic Poincar\'e map $\mathcal{P}: \mathbb{R}^2 \to \mathbb{R}^2$ sampled at integer multiples of the driving period $T = 2\pi/\Omega$.

The proposed architecture evaluates the dynamical state vertically across three distinct physical tiers, as conceptualized in Fig.~\ref{fig:concept}:

\subsection{Tier I: Microscopic Topology (Geometric Capacity)}
The foundational tier captures the microscopic geometric skeleton of the phase space, governed by the invariant manifolds of the primary saddle fixed point $z^*$ of the Poincar\'e map $\mathcal{P}$. The stable manifold $W^s(z^*)$ and unstable manifold $W^u(z^*)$ define global phase-space transport boundaries. A transversal intersection between $W^s(z^*)$ and $W^u(z^*)$ signifies a homoclinic tangency. By the Smale-Birkhoff theorem, this transversal intersection guarantees the existence of a Smale horseshoe, yielding a positive topological entropy ($h_T > 0$). This tier represents the \textit{informational capacity} or the geometric potential for complex dynamics, determined independently of phase-space contraction or thermodynamic energy balances.

\subsection{Tier II: Ergodic Measure (Informational Production)}
The intermediate tier evaluates the asymptotic physical measure and the actual informational production rate realized by the steady-state attractor. This is quantified by the Kolmogorov-Sinai (KS) metric entropy $h_{KS}$. According to Pesin's theorem \cite{Pesin1977}, for dissipative systems possessing an ergodic Sinai-Ruelle-Bowen (SRB) physical measure \cite{Eckmann1985}, the metric entropy is bounded by the sum of the positive Lyapunov exponents: $h_{KS} \le \sum_{\lambda_i>0} \lambda_i$. In our two-dimensional Poincar\'e map, the onset of physically observable chaos requires a positive maximal Lyapunov exponent ($\lambda_1 > 0$). Due to continuous phase-space volume contraction in dissipative systems, a geometric horseshoe (Tier I) may fail to attract a set of positive Lebesgue measure, rendering it a non-attracting transient chaotic repeller \cite{Grebogi1983}. In such regions, despite topological complexity ($h_T > 0$), the asymptotic measure collapses onto periodic sinks, yielding $h_{KS} = 0$.

\subsection{Tier III: Macroscopic Ensemble (Energetic Realization)}
The topmost macroscopic tier represents the coarse-grained statistical observables directly accessible in physical and engineering experiments, such as the steady-state kinetic variance $\sigma_x^2$ or macroscopic transport fluxes. This tier reflects the thermodynamic energy channel---specifically, the macroscopic power balance between external work injection and viscous dissipation. Macroscopic indicators are projections of the underlying high-dimensional manifold structure; they record the presence of energetic fluctuations but lack the topological resolution to distinguish whether these fluctuations originate from primary manifold tangles, higher-order saddle relays, or stochastic noise.

\subsection{Defining Structural Decoupling}
The traditional diagnostic assumption in applied nonlinear dynamics is that these three tiers synchronize: a homoclinic intersection (Tier I) induces a positive Lyapunov exponent (Tier II) and macroscopic variance (Tier III). However, our hierarchical evaluation reveals that this coupling breaks down systematically. An \textit{informational collapse} occurs when trajectories are caught in a non-attracting chaotic saddle (e.g., Point B in Fig.~\ref{fig:concept}). Here, the energy channel drives the system into a periodic sink (Tier III is phase-locked, Tier II yields $h_{KS}=0$), while the topological channel remains active (Tier I yields $h_T > 0$). Conversely, an \textit{informational relay} occurs when primary homoclinic intersections are severed, yet alternative higher-order saddles sustain macroscopic chaos (Point C). By mapping these structural decouplings across the $(\Omega, F)$ parameter space, we identify the physical origins of predictive failures in traditional geometric criteria. Crucially, this three-tier framework overcomes the structural limitations of single-layer diagnostics: conventional geometric evaluations (such as the Melnikov method) fail to determine attractor survivability (yielding false positives, as in Point B), while purely macroscopic statistical observables fail to detect the hidden higher-order manifold topologies that sustain turbulence after primary structures collapse (yielding false negatives, as in Point C).

%%%%%
\section{Structural Decoupling in Local Oscillations: The Duffing Paradigm}
\label{sec:duffing}

To substantiate the three-tier hierarchical architecture, we utilize the driven two-well Duffing oscillator (Eq.~\ref{eq:duffing}) with standard damping $k=0.1$. The state-space vector is defined by $\mathbf{z}(t) = (x(t), y(t))^T \in \mathbb{R}^2$ with $y = \dot{x}$. 

\subsection{High-Throughput Deterministic Manifold Binarization and Spectral Framework}
\label{sec:duffing_methods}

Evaluating the structural decoupling requires computing each tier independently without mutual contamination or numerical interpolation artifacts. Detailed numerical integration schemes, initial condition configurations to exclude multistability artifacts, and finite-time truncation validations for both the Duffing oscillator and the rotational pendulum are provided in the Appendix.

%%%% Figure 2 %%%%
\begin{figure*}[htbp]
\centering
\includegraphics[width=0.75\textwidth]{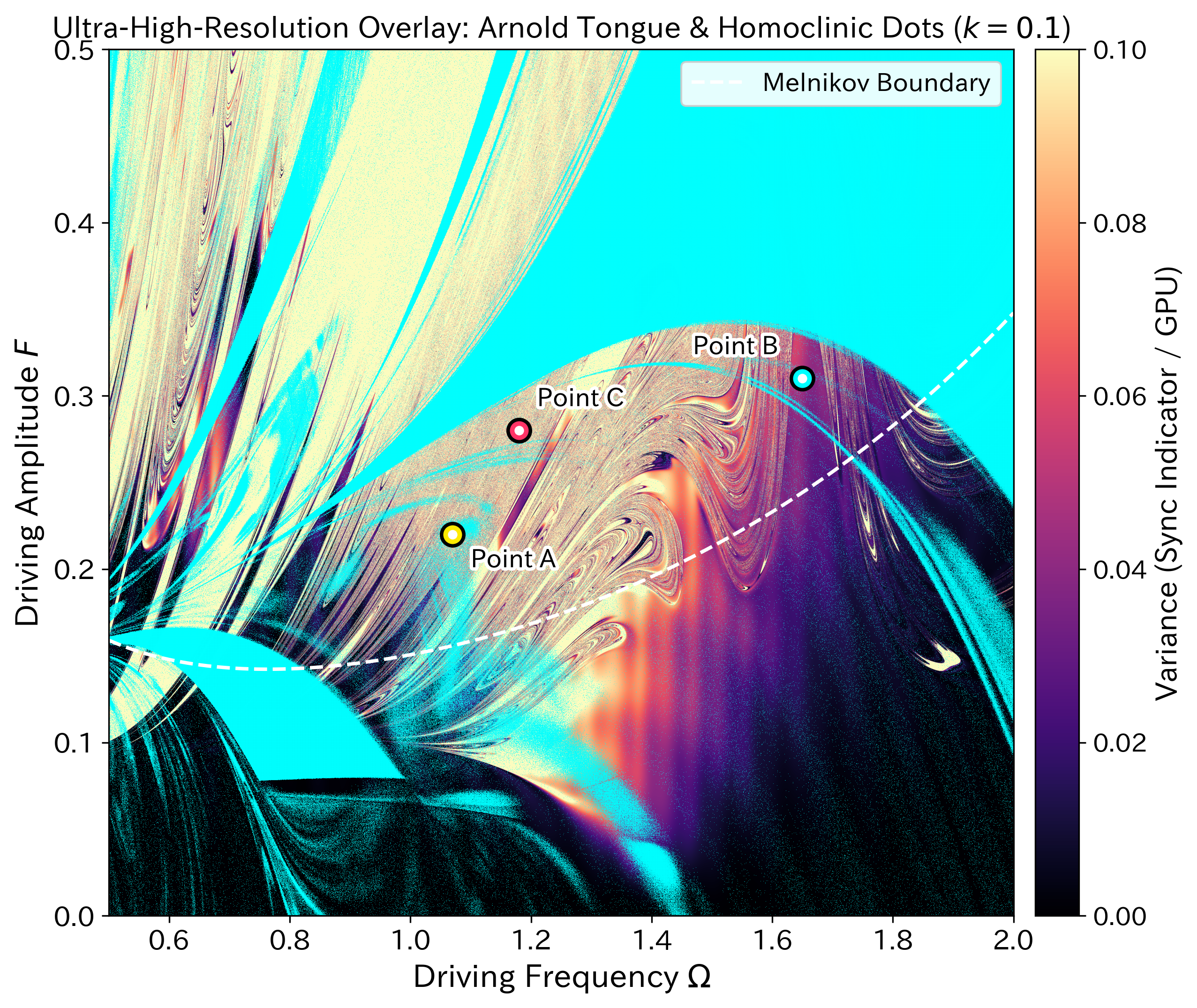}
\caption{\textbf{Global parameter space topology and macroscopic physical observability for the driven Duffing oscillator ($k = 0.1$).} 
The background colormap displays the steady-state kinetic variance $\sigma_x^2$ (Tier~III), delineating the infrastructure of nonlinear resonance tongues and large-scale transport. Deterministically evaluated primary transversal homoclinic intersections ($\Theta_{h_T} = 1.0$, Tier~I) are overlaid as semi-transparent cyan dots. The theoretical first-order Melnikov boundary is indicated by the white dashed curve. Regions above the Melnikov threshold exhibit dense microscopic tangles (cyan) while remaining embedded within non-fluctuating periodic attractors ($\sigma_x^2 = 0$). Curated representative coordinates are marked as Points A, B, and C. Note that the sharp, cliff-like boundaries bounding certain cyan domains (most visibly the abrupt vertical cutoff around $\Omega \approx 0.85, F \in [0.08, 0.17]$) are characteristic numerical truncation artifacts stemming from the finite integration horizon ($n = 8$ periods) of the manifold tracking algorithm, contrasting sharply with the smooth, continuous variation of the macroscopic variance.}
\label{fig:variance_overlay}
\end{figure*}

%%%% Figure 3 %%%%
\begin{figure*}[htbp]
\centering
\includegraphics[width=0.75\textwidth]{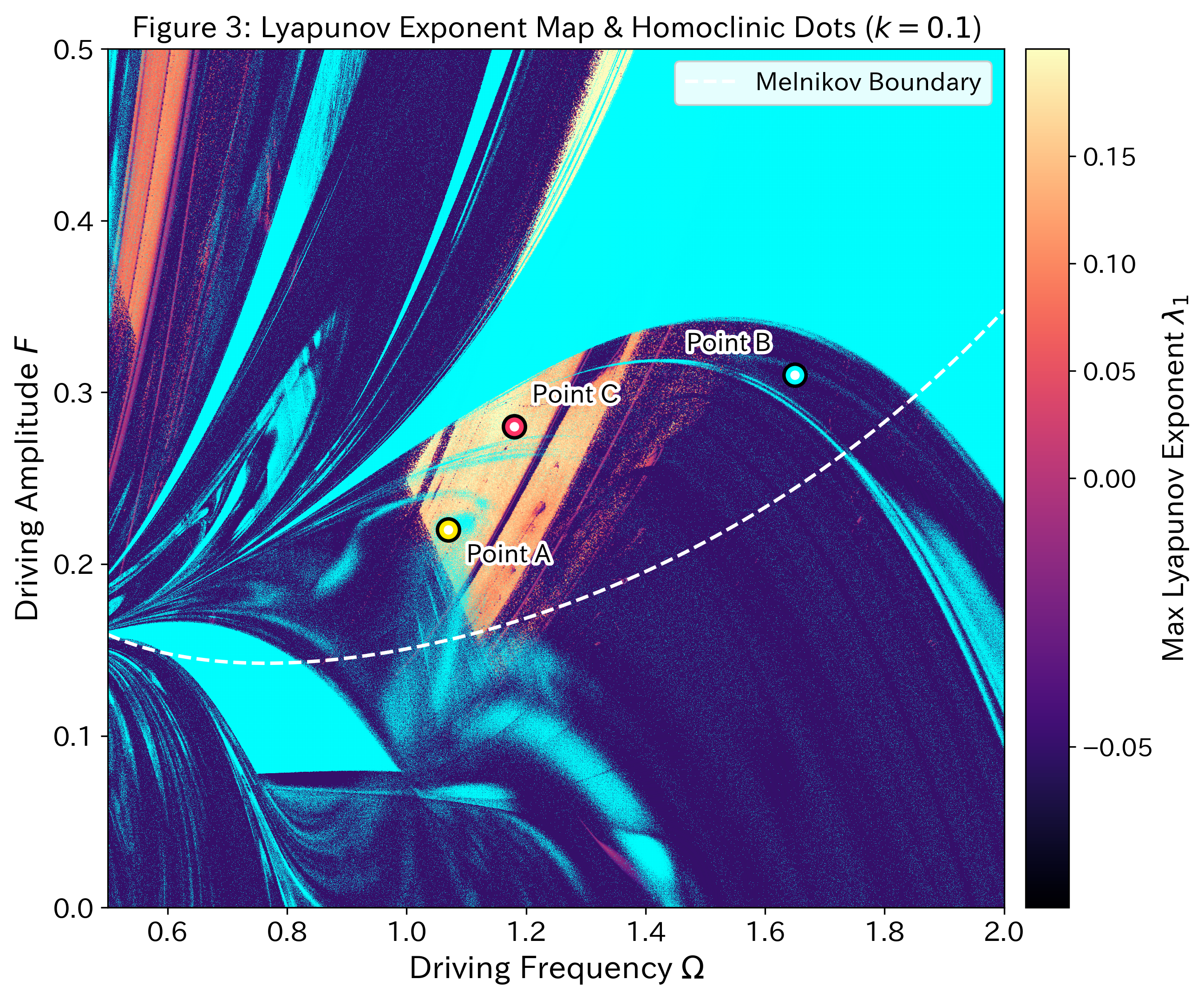}
\caption{\textbf{Direct visualization of entropic decoupling: Maximal Lyapunov exponent $\lambda_1$ overlaid with microscopic homoclinic intersections ($k = 0.1$).} 
The background colormap quantifies the maximal Lyapunov exponent $\lambda_1$ and asymptotic metric entropy $h_{KS}$ (Tier~II). Superimposed cyan dots denote active primary homoclinic crossings ($\Theta_{h_T} = 1.0$, Tier~I), and the white dashed curve indicates the Melnikov boundary. Curated representative coordinates are marked as Points A, B, and C. A comparison with Fig.~\ref{fig:variance_overlay} reveals: (i) non-attracting transient saddles where $h_T > 0$ (cyan) coexists with $\lambda_1 < 0$ (dark blue/purple), and (ii) higher-order saddle relays around $\Omega \in [1.1, 1.3]$ and $F \in [0.22, 0.32]$, where sustained ergodic chaos ($\lambda_1 > 0.10$, orange/yellow) persists despite the complete absence of primary homoclinic intersections. Note also that the abrupt, discontinuous termination of the cyan regions reflects the finite-time truncation of manifold growth, highlighting the necessity of cross-referencing with the continuous ergodic measure ($\lambda_1$).}
\label{fig:lyapunov_overlay}
\end{figure*}

\paragraph{Tier I (Microscopic Topology, $h_T$):}
For each parameter pair in the $(\Omega, F)$ domain, the primary hyperbolic saddle fixed point $\mathbf{z}^* = (x^*, y^*)^T$ of the Poincar\'e map $\mathcal{P}$ is dynamically tracked via a Newton--Raphson continuation scheme with a convergence tolerance of $\|\mathcal{P}(\mathbf{z}^*) - \mathbf{z}^*\| < 10^{-10}$. The unstable and stable eigenspaces are obtained from the local Jacobian, yielding the unit eigenvectors $\mathbf{v}_u$ and $\mathbf{v}_s$ associated with eigenvalues $|\mu_u| > 1$ and $|\mu_s| < 1$. Initial perturbation segments along the unstable branch are parameterized by $\mathbf{z}_0(\delta) = \mathbf{z}^* + \delta \mathbf{v}_u$ for $\delta \in [10^{-5}, 10^{-3}]$.

To evaluate transversal crossings deterministically, we deploy a GPU-accelerated ensemble engine executing an adaptive nonlinear density allocation. The evolved unstable manifold branch $W^u(\mathbf{z}^*)$ is generated by integrating an ensemble of 40 dynamically tracked points logarithmically distributed along the linear unstable eigenspace ($\delta \in [10^{-4}, 10^{-1}]$) across $n = 8$ Poincar\'e periods using a 4th-order Runge--Kutta scheme ($\Delta t = T / 128$). Transversal intersections with the stable manifold direction $\mathbf{v}_s$ are identified via the signed scalar distance function:
\begin{equation}
D(\delta) = \frac{(\mathcal{P}^8(\mathbf{z}_0(\delta)) - \mathbf{z}^*) \times \mathbf{v}_s}{\|\mathbf{v}_s\|}.
\label{eq:signed_distance}
\end{equation}
A deterministic binary flag $\Theta_{h_T}(\Omega, F)$ is assigned across the parameter grid:
\begin{equation}
\Theta_{h_T}(\Omega, F) = 
\begin{cases} 
1.0 \quad (h_T > 0), & \text{if } D(\delta) \text{ exhibits transverse sign reversals}, \\
0.0 \quad (h_T = 0), & \text{otherwise}.
\end{cases}
\label{eq:binary_flag}
\end{equation}
By the Smale--Birkhoff theorem, $\Theta_{h_T} = 1.0$ guarantees the existence of a Smale horseshoe ($h_T > 0$). In the global parameter maps (Figs.~\ref{fig:variance_overlay} and \ref{fig:lyapunov_overlay}), these points are rendered explicitly as semi-transparent cyan dots.

\paragraph{Computational Boundaries and Truncation Constraints of Manifold Tracking:}
While the GPU-accelerated binarization provides unprecedented parameter resolution, two fundamental limitations of invariant manifold computation must be emphasized. First, invariant manifolds are formally defined over asymptotic limits ($t \to \pm\infty$), whereas numerical tracking necessarily truncates the integration to a finite period ($n = 8$). While this capture window is sufficient to identify primary transverse crossings governing large-scale horseshoes, highly localized homoclinic tangles requiring longer transit times remain beyond finite-time detection. 

A prominent visual manifestation of this finite-time constraint appears directly in Figs.~\ref{fig:variance_overlay} and \ref{fig:lyapunov_overlay}, where the cyan domain exhibits abrupt, cliff-like boundaries---most visibly the sharp vertical cutoff around $\Omega \approx 0.85, F \in [0.08, 0.17]$, as well as the sudden termination of the high-frequency filament. These discontinuous boundaries do not represent sudden physical cessations of homoclinic tangency; rather, they mark the numerical threshold where the manifold arc length developed within $n = 8$ periods falls just short of intersecting the stable direction. Second, tracking is conditioned on the primary saddle fixed point $\mathbf{z}^*$ continued from the unforced center. Consequently, this geometric scan detects solely the topological entropy generated by the primary saddle's invariant manifolds ($h_T^{(\text{primary})} > 0$), leaving the topological contributions of bifurcated secondary subharmonic saddles unrecorded. As demonstrated below, these intrinsic geometric limitations of single-saddle, finite-time tracking provide the decisive justification for cross-auditing Tier~I with Tier~II.

\paragraph{Tier II (Ergodic Physical Measure, $h_{KS}$):}
The asymptotic rate of physical information production is computed via the maximal Lyapunov exponent $\lambda_1$ utilizing Benettin's orthonormalization algorithm \cite{Benettin1980}. Trajectories are integrated over $1.1 \times 10^5$ forcing periods; the initial $1.0 \times 10^4$ periods are discarded to eliminate transients, and $\lambda_1$ is accumulated over the subsequent $1.0 \times 10^5$ cycles. In accordance with Pesin's identity, the metric entropy is evaluated as:
\begin{equation}
h_{KS} = \max(0, \lambda_1).
\label{eq:pesin_entropy}
\end{equation}
The global distribution of $\lambda_1$ is mapped in Fig.~\ref{fig:lyapunov_overlay}, where $\lambda_1 \le 0$ indicates zero metric entropy ($h_{KS} = 0$), and positive values ($\lambda_1 > 0$) mark sustained ergodic chaos.

\paragraph{Tier III (Macroscopic Ensemble Statistics, $\sigma_x^2$):}
Macroscopic physical observability and kinetic energy dispersion are captured by the steady-state variance:
\begin{equation}
\sigma_x^2 = \lim_{N_p \to \infty} \frac{1}{N_p} \sum_{n=1}^{N_p} \left( x(n T) - \langle x \rangle \right)^2,
\label{eq:variance_def}
\end{equation}
accumulated over the asymptotic stroboscopic trajectory ($N_p = 5 \times 10^4$), where $\langle x \rangle$ is the mean position. For single-period phase-locked limit cycles, $\sigma_x^2 = 0$ identically, whereas chaotic inter-well transport yields a finite positive value ($\sigma_x^2 > 0$), as presented in the colormap of Fig.~\ref{fig:variance_overlay}.

\subsection{Global Decoupling Across the $(\Omega, F)$ Parameter Space}
\label{sec:duffing_global}

A comparative inspection of Figs.~\ref{fig:variance_overlay} and \ref{fig:lyapunov_overlay} over $(\Omega, F) \in [0.5, 2.0] \times [0.0, 0.5]$ reveals the structural decoupling operating between the three tiers. For analytical comparison, the first-order perturbative Melnikov threshold for the two-well Duffing oscillator:
\begin{equation}
F_{\text{Mel}}(\Omega) = \frac{4 k}{3 \pi \sqrt{2}} \cosh\left(\frac{\pi \Omega}{2}\right) \approx 0.0300 \cosh(1.5708 \Omega),
\label{eq:melnikov_formula}
\end{equation}
is overlaid as the white dashed curve in both figures. It is immediately apparent that while the underlying physical fields ($\sigma_x^2$ and $\lambda_1$) vary smoothly across the parameter plane, the Tier~I cyan regions exhibit sharp, unnatural cutoffs (such as the cliff-like vertical drop at $\Omega \approx 0.85$). These discontinuous edges are classical signatures of the finite-time numerical truncation ($n = 8$) discussed in Sec.~\ref{sec:duffing_methods}, demonstrating that relying solely on finite-horizon manifold tracking introduces artificial geometric boundaries that do not correspond to physical transitions in the system's observable state.

Classically, Eq.~(\ref{eq:melnikov_formula}) has been presumed to define the boundary of chaotic motion. However, our high-density overlay reveals systematic breakdowns of this assumption:

First, in the high-frequency domain ($\Omega \in [1.4, 2.0]$), a prominent arc-shaped filament of active homoclinic intersections (cyan dots, Tier~I) extends above the Melnikov line up to $F \approx 0.32$. However, Fig.~\ref{fig:lyapunov_overlay} confirms that the Lyapunov exponent throughout this entire cyan band remains strictly negative ($\lambda_1 < 0 \Rightarrow h_{KS} = 0$). Similarly, Fig.~\ref{fig:variance_overlay} demonstrates that the steady-state variance is zero ($\sigma_x^2 = 0$). This parameter domain provides evidence of non-attracting transient chaotic saddles: microscopic topological tangles ($h_T > 0$) exist, yet fail to generate an asymptotic ergodic measure or observable macroscopic fluctuation.

Second, in the central resonant region ($\Omega \in [1.0, 1.4]$), Fig.~\ref{fig:lyapunov_overlay} displays a robust chaotic sea ($\lambda_1 > 0.12$) accompanied by macroscopic variance ($\sigma_x^2 \approx 0.08 - 0.10$ in Fig.~\ref{fig:variance_overlay}). Crucially, within the parameter window $\Omega \in [1.10, 1.30]$ and $F \in [0.22, 0.32]$, there exists an extensive defect window where the cyan homoclinic dots are absent ($\Theta_{h_T} = 0.0$). In this regime, the primary saddle's invariant manifolds fail to intersect, yet macroscopically observable chaos is sustained.

\subsection{Microscopic Mechanisms at Representative Points A, B, and C}
\label{sec:duffing_cases}

The multi-tier behaviors manifested across Figs.~\ref{fig:variance_overlay} and \ref{fig:lyapunov_overlay} directly correspond to the three representative coordinates identified in the conceptual framework of Fig.~1:

\paragraph{Point A ($\Omega = 1.07, F = 0.22$ --- Congruent Ergodic Chaos):}
At Point A, all three tiers are synchronously active. We locate this point deep within the core of the primary chaotic resonance, safely separated from non-chaotic Arnold tongues. Microscopic manifold root-finding yields dense transversal intersections ($\Theta_{h_T} = 1.0$). Concurrently, Fig.~\ref{fig:lyapunov_overlay} indicates a robust positive maximal Lyapunov exponent ($\lambda_1 > 0.10$), and Fig.~\ref{fig:variance_overlay} confirms large-scale inter-well transport ($\sigma_x^2 \approx 0.08$). Here, the unstable manifold of the primary saddle folds back onto its stable manifold to construct an attracting chaotic set supporting a physical SRB measure. The informational channel directly feeds the macroscopic energetic channel.

\paragraph{Point B ($\Omega = 1.80, F = 0.28$ --- Non-Attracting Transient Saddle):}
Point B exemplifies the false-positive diagnosis of classical geometric criteria. Selected explicitly on the isolated homoclinic arc above the high-frequency Melnikov boundary, the parameter coordinate is populated by robust transversal intersections ($\Theta_{h_T} = 1.0$). However, the background of Fig.~\ref{fig:lyapunov_overlay} is strictly negative ($\lambda_1 \le -0.05 \Rightarrow h_{KS} = 0$), and Fig.~\ref{fig:variance_overlay} confirms a vanishing macroscopic variance ($\sigma_x^2 = 0$). 
In state space, the primary saddle generates a Cantor set of unstable chaotic orbits. However, phase-space dissipation continuously drains kinetic momentum, causing the unstable manifold branches to escape the tangled region and fall into the basin of attraction of coexisting period-1 sinks, as characteristic of boundary crises \cite{Grebogi1983}. The invariant chaotic set possesses zero Lebesgue measure; hence, macroscopic energy transport collapses into laminar motion while microscopic topological capacity remains active.

\paragraph{Point C ($\Omega = 1.18, F = 0.28$ --- Higher-Order Saddle Relay):}
Point C embodies the false-negative limitation of single-manifold indicators. This coordinate is deliberately selected at the geometric center of the macroscopic defect pocket ($\Omega \in [1.1, 1.3], F \in [0.22, 0.32]$) in Figs.~\ref{fig:variance_overlay} and \ref{fig:lyapunov_overlay}. Here, the primary saddle's invariant manifolds are completely detached ($\Theta_{h_T} = 0.0$). Conventional Melnikov analysis falsely classifies this state as regular. 
Nevertheless, both Fig.~\ref{fig:lyapunov_overlay} and Fig.~\ref{fig:variance_overlay} verify strong, sustained chaotic dynamics ($\lambda_1 > 0.10$, $\sigma_x^2 \approx 0.08$). Although the primary saddle's manifolds are strictly separated, the macroscopic SRB measure maintains a robust positive Lyapunov exponent. This coexistence logically dictates the activation of a higher-order saddle relay mechanism: secondary hyperbolic periodic saddles embedded in the inter-well region undergo mutual transverse intersections. In accordance with the Newhouse phenomenon \cite{Newhouse1979}, wild hyperbolic sets and dense subharmonic tangencies proliferate in the vicinity of homoclinic bifurcations, sustaining the chaotic transport channel. This secondary tangling maintains the ergodic mixing of the ensemble. Thus, sustained dissipative chaos can be preserved across parameter space through higher-order saddle relays, long after the primary homoclinic geometry is dismantled.

%%%%%%%
\section{Generalization to Global Rotational Transport: Driven Periodic-Potential Systems}
\label{sec:rotational}

To demonstrate that the hierarchical decoupling of informational and energetic channels represents a universal law of driven dissipative dynamics rather than a feature specific to polynomial multi-well potentials, we extend the three-tier framework to systems with periodic potentials. Specifically, we examine the driven, damped pendulum, which models macroscopic rotational transport across Josephson junctions and charge-density waves:
\begin{equation}
\ddot{\theta} + \beta \dot{\theta} + \sin\theta = F \cos(\Omega t),
\label{eq:pendulum_model}
\end{equation}
where $\theta \in \mathbb{S}^1$ denotes the macroscopic phase angle, $\beta$ is the viscous damping coefficient, $\Omega$ is the external driving frequency, and $F$ represents the excitation amplitude. Equation~(\ref{eq:pendulum_model}) corresponds directly to the canonical resistively shunted junction (RSJ) model of Stewart and McCumber governing macroscopic phase dynamics across Josephson junctions \cite{Stewart1968}. Throughout this section, the dissipation parameter is fixed to $\beta = 0.25$. 

The cylindrical state space $\mathbb{M} = \mathbb{S}^1 \times \mathbb{R}$ hosts two topologically distinct dynamical regimes: localized libration (oscillations trapped within a single potential well) and global rotation (running states continuously traversing the periodic potential barriers). The stroboscopic Poincar\'e map is defined on the cylinder as $\mathcal{P}_\theta: \mathbb{M} \to \mathbb{M}$, mapping $(\theta(t_0), \dot{\theta}(t_0)) \mapsto (\theta(t_0 + 2\pi/\Omega), \dot{\theta}(t_0 + 2\pi/\Omega)) \pmod{2\pi}$.

%%%% Figure 4 %%%%
\begin{figure*}[htbp]
\centering
\includegraphics[width=0.75\textwidth]{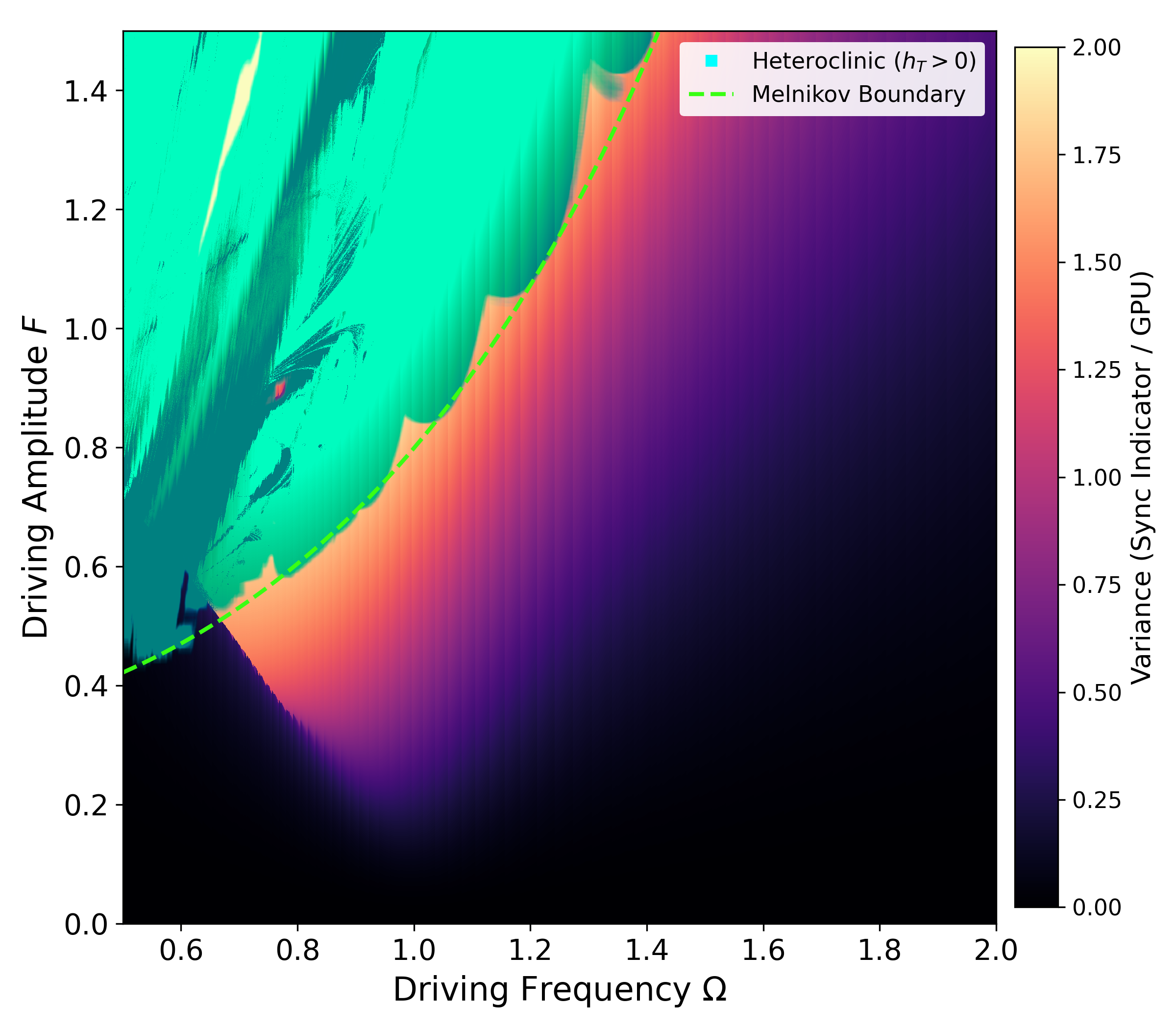}
\caption{\textbf{Global heteroclinic network and macroscopic transport variance in the driven periodic-potential system ($\beta = 0.25$).} 
The background colormap displays the steady-state kinetic variance (synchronization indicator, Tier~III) spanning from zero (localized librations, black) to large-scale rotational transport ($\sigma^2 \approx 2.0$, bright yellow/white). Overlayed cyan regions denote parameter domains with verified transversal heteroclinic intersections ($\Theta_{h_T} = 1.0$, Tier~I) emanating from the saddle states at $\theta = \pm \pi$. The green dashed line indicates the analytical first-order Melnikov boundary for rotational separatrix crossing. The heteroclinic onset displays scalloped resonance boundaries, confirming the geometric opening of global phase-slipping channels across adjacent potential wells.}
\label{fig:pendulum_variance_overlay}
\end{figure*}

%%%% Figure 5 %%%%
\begin{figure*}[htbp]
\centering
\includegraphics[width=0.75\textwidth]{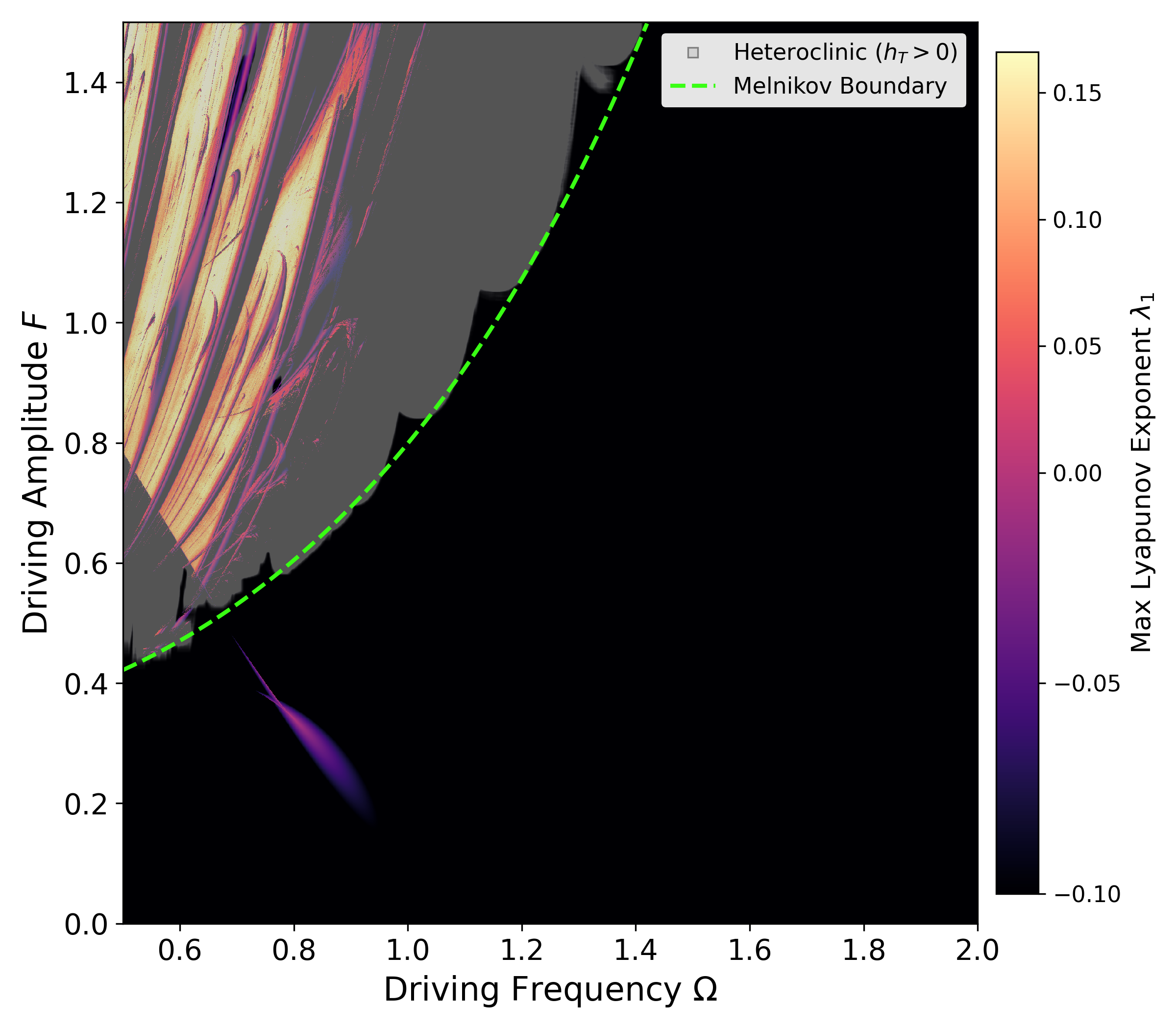}
\caption{\textbf{Direct proof of informational collapse in global rotational transport ($\beta = 0.25$).} 
The background colormap displays the maximal Lyapunov exponent $\lambda_1$ (Tier~II, ergodic measure). Overlayed semi-transparent gray shading marks the domain of active microscopic heteroclinic tangles ($\Theta_{h_T} = 1.0$, Tier~I), bounded below by the analytical Melnikov line (green dashed curve). A dynamic decoupling is revealed: across the vast majority of the heteroclinic domain, the maximal Lyapunov exponent remains non-positive ($\lambda_1 \le 0 \Rightarrow h_{KS} = 0$), corresponding to phase-locked, non-chaotic Shapiro transport. Sustained ergodic chaotic transport ($\lambda_1 > 0$, bright orange-yellow filaments) is confined to narrow resonance-overlap channels at low driving frequencies ($\Omega \in [0.5, 1.0]$).}
\label{fig:pendulum_lyapunov_overlay}
\end{figure*}

\subsection{Macroscopic Observables and Transport Energetics (Tier III)}
\label{sec:rot_observables}

In rotational systems, the energetic channel (Tier~III) governs both the macroscopic fluctuation variance and the directed transport flux. The steady-state kinetic variance $\sigma^2$ quantifies the overall dynamic dispersion, as mapped in Fig.~\ref{fig:pendulum_variance_overlay}. 

Complementing the variance, the net macroscopic transport flux is quantified by the time-averaged rotational velocity:
\begin{equation}
\langle \dot{\theta} \rangle = \lim_{T \to \infty} \frac{1}{T} \int_0^T \dot{\theta}(t) \, dt.
\label{eq:dc_velocity}
\end{equation}
The thermodynamic energy balance requires that external power injection balances viscous heat dissipation over asymptotic cycles:
\begin{equation}
\langle \mathcal{W}_{\text{in}} \rangle = \lim_{T \to \infty} \frac{1}{T} \int_0^T F \cos(\Omega t) \, \dot{\theta}(t) \, dt = \beta \langle \dot{\theta}^2 \rangle = \langle \mathcal{Q}_{\text{diss}} \rangle.
\label{eq:power_balance}
\end{equation}
When the system entrains into periodic running states, $\langle \dot{\theta} \rangle$ locks onto discrete plateaus known as Shapiro steps \cite{Shapiro1963}, satisfying $\langle \dot{\theta} \rangle = \frac{p}{q} \Omega$ with $p, q \in \mathbb{Z}$. Across these steps, kinetic variance remains bounded while energy flows continuously through the system.

\subsection{Microscopic Heteroclinic Topology vs. Melnikov Bounds (Tier I)}
\label{sec:rot_topology}

The microscopic geometric skeleton (Tier~I) is governed by the invariant manifolds of the unstable hyperbolic saddle points located at $\mathbf{z}_s = (\pm \pi, 0)$. In the conservative limit ($\beta = 0, F = 0$), separatrices connect adjacent saddles, forming unperturbed heteroclinic loops that demarcate the boundary between libration and rotation. Under finite dissipative driving ($\beta = 0.25, F > 0$), first-order Melnikov perturbation theory predicts that stable ($W^s$) and unstable ($W^u$) manifolds intersect transversally across adjacent spatial potential wells when:
\begin{equation}
F \ge F_{\text{Mel}}^{(\text{rot})}(\Omega) = \frac{4 \beta}{\pi} \cosh\left(\frac{\pi \Omega}{2}\right) \approx 0.3183 \cosh(1.5708 \Omega).
\label{eq:rot_melnikov}
\end{equation}
Equation~(\ref{eq:rot_melnikov}) is overlaid as the green dashed curve in Figs.~\ref{fig:pendulum_variance_overlay} and \ref{fig:pendulum_lyapunov_overlay}.

Using GPU-accelerated root-finding across the $(\Omega, F)$ plane, we deterministically identify active transversal heteroclinic crossings ($\Theta_{h_T} = 1.0$, rendered in cyan in Fig.~\ref{fig:pendulum_variance_overlay} and gray in Fig.~\ref{fig:pendulum_lyapunov_overlay}). Geometrically, heteroclinic tangency eliminates the final rotational Kolmogorov-Arnold-Moser (KAM) barriers, establishing a global Smale horseshoe network across spatial periods and guaranteeing positive topological entropy ($h_T > 0$). The numerical boundary exhibits scalloped resonance tongues departing from the smooth Melnikov approximation, reflecting non-perturbative subharmonic resonances.

\subsection{Hierarchical Decoupling in Shapiro Steps and Chaotic Transport}
\label{sec:rot_decoupling}

A simultaneous comparison between Tier~III (Fig.~\ref{fig:pendulum_variance_overlay}) and Tier~II (Fig.~\ref{fig:pendulum_lyapunov_overlay}) provides clear evidence of hierarchical decoupling in global transport:

\paragraph{Informational Collapse in Coherent Shapiro Transport ($h_T > 0, h_{KS} = 0, \sigma^2 > 0$):}
Across the broad region of the heteroclinic domain above the Melnikov threshold (shaded gray in Fig.~\ref{fig:pendulum_lyapunov_overlay}), the maximal Lyapunov exponent is strictly non-positive ($\lambda_1 \le 0$). In this extensive region, the Kolmogorov--Sinai metric entropy collapses to zero ($h_{KS} = 0$). 
Simultaneously, Fig.~\ref{fig:pendulum_variance_overlay} confirms that macroscopic transport is operational: trajectories exhibit finite variance ($\sigma^2 > 0$) and locked non-zero transport velocities ($\langle \dot{\theta} \rangle = n \Omega$). 
This state represents the global rotational manifestation of Point~B from the Duffing system (non-attracting transient saddle). Although the heteroclinic network provides the topological permission for phase slipping across potential barriers, dissipative volume contraction along the running trajectory forces all ensembles into stable, zero-entropy Shapiro limit cycles. Coherent macroscopic energy transport persists while microscopic information generation is entirely extinguished.

\paragraph{Ergodic Chaotic Phase Slipping ($h_T > 0, h_{KS} > 0$):}
Sustained chaotic transport ($\lambda_1 > 0$, bright orange/yellow filaments in Fig.~\ref{fig:pendulum_lyapunov_overlay}) occurs within narrow resonance-overlap channels concentrated in the low-frequency domain $\Omega \in [0.5, 1.0]$ and along the destabilization margins of major Shapiro steps. In these channels, all three tiers are synchronously active: the topological horseshoe becomes an attracting chaotic set supporting an asymptotic SRB measure ($h_{KS} > 0$), resulting in intermittent phase slipping, broadband power spectra, and non-quantized diffusive flux.

These numerical observations reveal three fundamental dynamical principles governing non-equilibrium rotational transport. First, transversal heteroclinic intersection operates strictly as a geometric connectivity threshold; it provides the topological prerequisite for phase slips across adjacent potential barriers, but exerts no control over whether the resulting asymptotic state settles into coherent limit-cycle rotation or diffusive chaotic mixing. Second, macroscopic transport flux ($\langle \dot{\theta} \rangle > 0$) and energetic variance ($\sigma^2 > 0$) remain entirely independent of informational entropy production. Coherent directed currents can be sustained indefinitely in the complete absence of Kolmogorov--Sinai metric entropy ($h_{KS} = 0$). Finally, this operational independence demonstrates that the structural decoupling between informational and energetic channels is universal, governing both bounded multi-well oscillators and unbounded periodic-potential transport.

%%%%%
\section{Discussion: Physical Mechanisms and Theoretical Implications}
\label{sec:discussion}

The systematic evidence gathered across local double-well oscillations (Figs.~\ref{fig:variance_overlay} and \ref{fig:lyapunov_overlay}) and global rotational transport (Figs.~\ref{fig:pendulum_variance_overlay} and \ref{fig:pendulum_lyapunov_overlay}) demonstrates that the three-tier hierarchical architecture addresses discrepancies between theoretical and empirical findings in nonlinear mechanics.

\subsection{Resolving the Classical Divide: Topological Capacity vs. Ergodic Realization}
\label{sec:disc_divide}

For decades, non-conservative dynamical systems have been investigated through two largely disconnected conceptual schools. On one side stands the geometric and topological paradigm, rooted in the Smale--Birkhoff theorem and Melnikov perturbation theory, which equates deterministic chaos with the transversal intersection of invariant manifolds ($h_T > 0$). On the other side stands the observable and statistical paradigm, founded upon smooth ergodic theory and Lyapunov characteristic spectra, which identifies chaos solely through the asymptotic emergence of an invariant physical measure ($h_{KS} > 0$) accompanied by macroscopic broadband fluctuations ($\sigma^2 > 0$).

Our three-tier mapping proves that treating these paradigms as interchangeable is inaccurate in dissipative mechanics. Transversal manifold intersections establish solely the \textit{topological capacity} for complex dynamics—guaranteeing the existence of a non-empty, invariant Smale horseshoe. However, because dissipative phase-space contraction constantly contracts Lebesgue volume, the physical observability of this horseshoe depends entirely on its stability and measure-theoretic embedding. 

As demonstrated by Point~B in the Duffing system and the Shapiro step domain in the pendulum system, the horseshoe frequently manifests as a non-attracting chaotic saddle. Trajectories initialized along the manifold tangle disperse transiently across the saddle before being irreversibly drawn into periodic limit cycles. Consequently, the asymptotic SRB measure concentrates entirely on zero-entropy sinks, yielding strictly $h_{KS} = 0$ despite $h_T > 0$. Microscopic topological intersection is therefore fundamentally a criterion for \textit{transient chaos} \cite{Lai2011}, not sustained, observable chaos. The classical Melnikov boundary must be reinterpreted: it delineates the threshold where the capacity for topological complexity is born, rather than the ignition boundary of macroscopic physical turbulence.

\subsection{Orthogonality of Energetic Transport and Informational Entropy}
\label{sec:disc_orthogonality}

A foundational insight of this work is the structural orthogonality between the \textit{energetic channel} (Tier~III) and the \textit{informational channel} (Tiers~I and II). 

In the literature, large macroscopic transport or high dissipation has often been conflated with turbulent, chaotic mixing. The global rotational results in Sec.~\ref{sec:rotational} disprove this assumption. Within stable Shapiro resonance steps (Figs.~\ref{fig:pendulum_variance_overlay} and \ref{fig:pendulum_lyapunov_overlay}), this orthogonality becomes physically transparent. The energetic channel experiences robust sustained throughput: continuous external power injection compensates viscous friction according to the time-averaged balance $\langle \mathcal{W}_{\text{in}} \rangle = \beta \langle \dot{\theta}^2 \rangle > 0$, sustaining a macroscopic quantized flux $\langle \dot{\theta} \rangle = n \Omega$. Simultaneously, however, the informational channel is completely extinguished. Strong phase-space contraction along the running trajectory suppresses all localized divergence, driving the maximal Lyapunov exponent into negative territory ($\lambda_1 < 0$) and strictly quenching the metric entropy to zero ($h_{KS} = 0$).
The opening of heteroclinic channels ($h_T > 0$) is geometrically indispensable to permit phase-slip transitions across periodic barriers. However, once the barrier is traversed, the asymptotic dynamics within the energetic conduit can be completely coherent and regular. Energy transport does not require information generation; macroscopic flux can be sustained with zero Kolmogorov--Sinai entropy.

\subsection{Higher-Order Saddle Networks and the Methodological Limitations of Homoclinic Diagnostics}
\label{sec:disc_relays}

The computational discovery of Point~C in Fig.~\ref{fig:lyapunov_overlay} exposes a deep methodological bottleneck in conventional homoclinic and heteroclinic diagnostics. Historically, analytical Melnikov methods and numerical manifold algorithms have relied almost exclusively on tracking the invariant manifolds of the \textit{primary} saddle. When these manifolds detach ($\Theta_{h_T} = 0$), classical geometric criteria inevitably classify the dynamics as regular. However, our findings uncover two structural failure modes inherent to single-manifold evaluations in dissipative mechanics.

\subsubsection{Attractor Embedding and the False-Positive Dilemma}
The first fundamental limitation concerns the inability of geometric intersections to assess attractor stability. Transversal crossings between $W^u$ and $W^s$ confirm the topological existence of a Smale horseshoe, yet provide no information regarding whether the resulting invariant hyperbolic set is asymptotically attracting or transient. Under finite dissipative volume contraction, the primary horseshoe routinely degenerates into a non-retaining chaotic saddle, as manifested at Point~B. Consequently, topological intersection alone fails as a standalone criterion for observable turbulence.

\subsubsection{Subharmonic Cascades and the False-Negative Dilemma}
The second limitation arises from the blindness of single-saddle diagnostics to secondary and subharmonic manifold relays. In driven nonlinear oscillators, harmonic forcing generates an extensive hierarchy of secondary periodic saddles through cascades of period-doubling and saddle-node bifurcations, consistent with the proliferation of wild hyperbolic sets near homoclinic tangencies \cite{Newhouse1979}. Even after the primary saddle's invariant manifolds pull apart to form an apparent gap, the invariant manifolds of coexisting secondary saddles frequently maintain dense transversal tangles. The asymptotic physical measure is supported by this secondary manifold network, sustaining ergodic chaos ($h_{KS} > 0, \sigma_x^2 > 0$) directly across the primary defect pocket.

Computing homoclinic or heteroclinic crossings of a single saddle therefore provides neither a necessary nor a sufficient condition for dissipative chaos. Recognizing these limitations provides the decisive justification for the three-tier architecture: Tier~I establishes geometric connectivity, while Tiers~II and III certify the measure-theoretic and energetic survival of the motion. Practically, recognizing this higher-order relay mechanism enables the diagnosis of ``unexplained'' persistent vibrations or noise in engineering systems, specifically in parameter regimes where theoretical analyses of the primary saddle incorrectly predict a safe, regular state.

\subsection{Broader Scope: Multi-Degree-of-Freedom Extensions and Stochastic Vulnerability}
\label{sec:disc_extensions}

While demonstrated here on benchmark non-autonomous systems, the conceptual architecture and diagnostic methodology established in this study provide a foundation for broader physical domains.

\subsubsection{Coupled Dissipative Systems and Collective Transport}
The deterministic multi-tier methodology established in this study provides a natural foundation for analyzing multi-degree-of-freedom coupled dissipative networks. In spatially distributed mechanical systems---such as the nonlinearly coupled magneto-elastic beam arrays investigated by Hikihara \textit{et al.}~\cite{Hikihara2001}---macroscopic wave propagation is governed by the wandering of trajectories among coexisting saddle states through heteroclinic manifold intersections. 

Applying the three-tier framework to such coupled systems offers the strong potential to resolve the interplay between local chaos and collective energetic transport. Specifically, it enables a systematic investigation into whether macroscopic wave propagation and coherent energetic flux can be sustained across an oscillator lattice while local Kolmogorov--Sinai metric entropy remains spatially confined, and conversely, how the degeneration of local manifolds into non-attracting transient saddles can dictate the vulnerability of the entire network to noise-induced desynchronization and spatio-temporal phase defects.

\subsubsection{Stochastic Susceptibility and Noise-Induced Triggering}
In practical physical environments, non-attracting chaotic saddles are highly vulnerable to ambient fluctuations. As established in the seminal studies of noise-induced crises \cite{Sommerer1991}, thermal or external noise readily kicks trajectories out of narrow periodic attraction basins back into the surrounding transient saddle tangle, igniting noise-induced intermittent chaos. Similarly, in macroscopic quantum systems such as rf-biased Josephson junctions, such transient topological tangles drastically lower the activation barrier for thermally induced phase slips and loss of synchronization \cite{Kautz1985}. 

Mapping the underlying transient topological capacity ($h_T > 0, h_{KS} = 0$) therefore serves as an intrinsic sensitivity map for noise vulnerability. In engineering contexts, regions traditionally considered stable and safe (e.g., Point B) can now be systematically audited as hidden reservoirs of instability, where even microscopic perturbations can trigger sudden loss of synchronization or catastrophic phase slips.

Finally, while the present work establishes the phenomenological decoupling of these three tiers across continuous parameter spaces, formalizing the exact quantitative bridges connecting escape rates on transient saddles, turnstile lobe flux, and deterministic diffusion represents an important direction for future theoretical investigation.

\section{Conclusion}
\label{sec:conclusion}

In this study, we have addressed a long-standing structural disconnect in nonlinear dynamics: the discrepancy between abstract geometric criteria of chaos and macroscopically observable physical dynamics in driven dissipative systems. Recognizing that phase-space volume contraction inherently decouples information-theoretic entropy from thermodynamic energy balances, we formulated a unified three-tier hierarchical architecture. This framework explicitly bridges the microscopic topological capacity governed by invariant manifold tangles (Tier~I, $h_T$), the asymptotic physical measure quantified by Kolmogorov--Sinai metric entropy (Tier~II, $h_{KS}$), and the coarse-grained macroscopic ensemble statistics accessible in experiments (Tier~III, $\sigma_x^2$ or $\langle \dot{\theta} \rangle$).

Deploying high-throughput deterministic binarization and spectral maps to the bistable Duffing oscillator uncovered the dynamical origins of classical predictive failures. We proved that the false positives of the Melnikov criterion arise from non-attracting transient chaotic saddles, exemplified by Point~B, where dense geometric horseshoes collapse into zero-entropy periodic sinks and extinguish observable fluctuations. Conversely, we resolved the false-negative puzzle at Point~C by uncovering a higher-order saddle relay mechanism: macroscopic chaos persists across primary homoclinic defect windows through the cooperative tangling of bifurcated subharmonic saddle manifolds, exposing the fundamental inadequacy of monitoring only the primary saddle.

Generalizing the architecture to periodic-potential rotational dynamics established the operational orthogonality of macroscopic transport and information generation. Within resonant Shapiro steps, global heteroclinic networks open the spatial channels required for phase slipping, yet phase-space contraction quenches the Lyapunov spectrum into negative territory. Consequently, directed macroscopic energy currents operate at maximum thermodynamic throughput with strictly zero Kolmogorov--Sinai entropy, demonstrating that non-equilibrium transport flux does not require turbulent mixing.

Ultimately, evaluating dissipative dynamical systems vertically across all three tiers provides profound diagnostic power for applied physics and engineering. It allows practitioners to identify seemingly stable periodic regimes as latent reservoirs of noise-induced instability, while correctly diagnosing anomalous persistent vibrations through higher-order topological cascades. By structurally distinguishing geometric permission from asymptotic physical realization, this three-tier architecture establishes a rigorous bridge between geometric mechanics and non-equilibrium statistical physics.

%\section*{Data Availability}
%The data that support the findings of this study are available from the corresponding author upon reasonable request.

\appendix
\renewcommand{\thesubsection}{A.\arabic{subsection}}

\section*{Appendix: Detailed Numerical Settings}
\addcontentsline{toc}{section}{Appendix: Detailed Numerical Protocols and Validations}
This appendix provides detailed descriptions of the numerical integration schemes, initial condition configurations, and finite-time truncation validations employed to map the three-tier hierarchical architecture in both the driven two-well Duffing oscillator (Section~\ref{sec:duffing}) and the driven periodic-potential rotational system (Section~\ref{sec:rotational}). These deterministic setups are specifically designed to isolate pure asymptotic physical measures and global topological invariant manifolds, rigorously excluding artifacts such as multistability hysteresis and finite-time integration leakage.

\subsection{Local Oscillations: The Driven Duffing Oscillator}
The macroscopic physical fields in the $(\Omega, F)$ parameter space are generated through ultra-high-resolution parallel scans $(5000 \times 5000 \text{ grid points})$.

\subsubsection{Tier III: Macroscopic Variance and Initialization}
To map the steady-state kinetic variance $\sigma_x^2$ without introducing continuation hysteresis (i.e., artificial dragging of coexisting attractors across parameter sweeps), all independent integration threads are initiated from a uniform, mathematically identical \textit{quiescent state}:
\begin{equation}
    (x_0, \dot{x}_0) = (0.1, 0.0).
\end{equation}
This specific coordinate intentionally offsets trajectories slightly from the central potential barrier maximum $(0, 0)$, allowing the deterministic dissipation field to guide the orbit naturally into its dominant asymptotic basin of attraction under the application of the external harmonic forcing $F \cos(\Omega t)$. By discarding the initial transients, this pure re-initialization guarantees that parameter domains exhibiting $\sigma_x^2 = 0$ (such as Point B) genuinely represent stable non-fluctuating periodic attractors, rather than artifacts of initial condition selection.

\subsubsection{Tier I: Primary Saddle Tracking and Truncation Validation}
The geometric extraction of transversal homoclinic tangencies $(\Theta_{h_T} = 1.0)$ requires extreme care in isolating the \textit{primary} unstable saddle from an exponentially growing number of subharmonic orbits generated near bifurcations (such as those governing the higher-order saddle relay at Point C).

\paragraph{Newton--Raphson Seeding:} To ensure that only the primary saddle continues to be tracked, the local Newton--Raphson root-finding algorithm is systematically seeded at a localized coordinate near the unforced saddle:
\begin{equation}
    (x_{\text{init}}, \dot{x}_{\text{init}}) = (-0.15, 0.02).
\end{equation}
This fixed seed structurally prevents the root-finder from diverging and mistakenly tracking secondary hyperbolic sets that are born through period-doubling cascades deep within the resonance tongues.

\paragraph{Finite-Time Integration Horizon ($n=8$):}
For the Duffing system operating under the fixed dissipation coefficient $k=0.1$, the absolute magnitude of the unstable Floquet multiplier ensures that the unstable invariant manifold arc length stretches exponentially by a factor exceeding $\sim 10^6$ over a horizon of 8 forcing periods (e.g., stretching by a factor of $\approx 2.4 \times 10^{10}$ even at the high-frequency limit $\Omega = 2.0$, and exceeding $10^{20}$ within the primary resonance). Evaluating an ensemble of 40 dynamically tracked points logarithmically distributed along the linear unstable eigenspace ($\delta \in [10^{-4}, 10^{-1}]$) out to $n = 8$ periods provides complete macroscopic phase-space coverage across the bounded potential wells. Crucially, resolving the primary global Smale horseshoes up to this numerical threshold is mathematically sufficient. While extending the integration horizon to $n=15$ or $20$ would marginally elongate the trailing tails of the manifold tangles (slightly softening the sharp truncation boundary observed in the high-frequency parameter map), doing so exacerbates exponential accumulation of numerical precision errors without altering the fundamental physical mechanism: the eventual, inevitable collapse of the transient chaotic saddle into surrounding periodic sinks.

\subsection{Global Rotational Transport: Driven Periodic-Potential System}
For the driven damped pendulum model governing macroscopic phase dynamics across Josephson junctions, the integration protocols are similarly engineered to decouple localized microscopic topology from global steady-state transport.

\subsubsection{Tier III: Energetic Transport and Transient Elimination}
Mapping the macroscopic transport flux (time-averaged rotational velocity $\langle \dot{\theta} \rangle$) and steady-state variance $\sigma^2$ across the $2000 \times 2000$ parameter grid requires strict elimination of rotational transients, particularly in the vicinity of global heteroclinic bifurcations.
Analogous to the local Duffing model, all trajectories are seeded from the uniform quiescent state:
\begin{equation}
    (\theta_0, \dot{\theta}_0) = (0, 0).
\end{equation}
To definitively purge the extremely long-lived transients characteristic of weakly damped $(\beta = 0.25)$ non-equilibrium systems drifting toward phase-locked Shapiro steps, trajectories are deterministically integrated for a massive initial horizon of $800$ forcing periods before any data is recorded. The subsequent $200$ periods are then utilized to accumulate the physical statistics $\sigma^2$ and $\langle \dot{\theta} \rangle$. This structural design confirms that the observed transport fields are purely asymptotic physical measures.

\subsubsection{Tier I: Heteroclinic Manifold Tracking}
The microscopic geometric skeleton dictating global phase slips relies on transversal heteroclinic crossings between adjacent unstable hyperbolic saddles.
\paragraph{Saddle Seed and Integration Horizon:} 
The Newton--Raphson root-finder is explicitly seeded near the primary rotational barriers at:
\begin{equation}
    (\theta_{\text{init}}, \dot{\theta}_{\text{init}}) = (\pm \pi, 0).
\end{equation}
Due to the extended spatial domain required to successfully cross from one periodic potential well to the adjacent well, the transversal intersection scan is executed over an extended numerical horizon of $20$ forcing periods. This ensures that the generated invariant manifold networks possess sufficient arc length to unambiguously confirm the opening of global topological transport channels (Smale horseshoes across spatial periods) independent of the subsequent informational collapse into zero-entropy Shapiro limit cycles.

\end{document}